\documentclass[11pt,a4paper]{article}
\pdfoutput=1

\usepackage{jcappub}
\usepackage[latin1]{inputenc}
\usepackage{graphicx}
\usepackage{epsfig}
\usepackage{subfigure}
\usepackage{color}
\def\beq{\begin{equation}}
\def\eeq{\end{equation}}
\def\baq{\begin{eqnarray}}
\def\eaq{\end{eqnarray}}

\newcommand{\be}{\begin{equation}} % only untightened
\newcommand{\ee}{\end{equation}}
\newcommand{\bea}{\begin{eqnarray}} % only untightened
\newcommand{\eea}{\end{eqnarray}}

\newcommand{\bmp}{\noindent\begin{minipage}{16cm}}
\newcommand{\emp}{\end{minipage}\vskip 7mm} % 7mm untightened
\def\lsim{\mathrel{\raise.3ex\hbox{$<$\kern-.75em\lower1ex\hbox{$\sim$}}}}
\def\gsim{\mathrel{\raise.3ex\hbox{$>$\kern-.75em\lower1ex\hbox{$\sim$}}}}

\newcommand{\intron}[1]{}%{#1}

\subheader{\small{Preprint: HIP-2014-11/TH}}

\title{Standard Model with a real singlet scalar and inflation}

\author{Kari Enqvist,}
\author{Sami Nurmi,}
\author{Tommi Tenkanen and}
\author{Kimmo Tuominen}

\affiliation{University of Helsinki and Helsinki Institute of Physics, \\
                      P.O.~Box 64, FI-00014, Helsinki, Finland}
\emailAdd{kari.enqvist@helsinki.fi}
\emailAdd{sami.nurmi@helsinki.fi}
\emailAdd{tommi.tenkanen@helsinki.fi}
\emailAdd{kimmo.i.tuominen@helsinki.fi}

\abstract{We study the post-inflationary dynamics of the Standard Model Higgs and a real singlet scalar $s$, coupled together through a renormalizable coupling $\lambda_{sh}h^2s^2$, in a $Z_2$ symmetric model that may explain the observed dark matter abundance and/or the origin of baryon asymmetry. The initial values for the Higgs and $s$ condensates are given by inflationary fluctuations, and we follow their dissipation and relaxation to the low energy vacua. We find that both the lowest order perturbative and the non-perturbative decays are blocked by thermal effects and large background fields and that the condensates decay by two-loop thermal effects. Assuming instant reheating at $T=10^{16}$ GeV, the characteristic temperature for the Higgs condensate thermalization is found to be $T_h \sim 10^{14}$ GeV, whereas $s$ thermalizes typically around $T_s \sim 10^{6}$ GeV. By that time, the amplitude of the singlet is driven very close to the vacuum value by the expansion of the universe, unless the portal coupling takes a value $\lambda_{sh}\lesssim 10^{-7}$ and the singlet $s$ never thermalizes. With these values of the coupling, it is possible to slowly produce a sizeable fraction of the observed dark matter abundance via singlet condensate fragmentation and thermal Higgs scattering. Physics also below the electroweak scale can therefore be affected by the non-vacuum initial conditions generated by inflation. 
}


\begin{document}
\maketitle
%%%%%%%%%%%%%%%%%%%%%%%%%%%%%%%%%%%%%%%%%%%%%%%%%%%%%%%%%%%%%%%%%%%%%%%%%%%%%%%%%%%%%%%%%%%%%%%%%%%%

%%%%%%%%%%%%%%%%%%%%%%%%%%%%%%%%%%%%%%%%%%%%%%%%%%%%%%%%%%%%%%%%%%%%%%%%%%%%%%%%%%%%%%%%%%%%%%%%%%%%
%%%%%%%%%%%%%%%%%%%%%%%%%%%%%%%%%%%%%%%%%%%%%%%%%%%%%%%%%%%%%%%%%%%%%%%%%%%%%%%%%%%%%%%%%%%%%%%%%%%%
%
\section{Introduction}
The discovery of the Higgs boson \cite{Aad:2012tfa,Chatrchyan:2012ufa} has provided a stringent test for beyond the Standard Model (BSM)
scenarios \cite{Carmi:2012yp,Espinosa:2012ir,Giardino:2012ww,Alanne:2013dra}.
Nevertheless traditional BSM paradigms, such as supersymmetry and technicolor, still remain as viable alternatives for a more fundamental theory underlying the electroweak sector.
The LHC data seem to indicate that the scalar sector underlying the electroweak symmetry breaking is very close to the simplest alternative, i.e. the SM
itself ~\cite{ATLAS-CONF-2013-012,CMS-PAS-HIG-13-001,ATLAS-CONF-2013-030,ATLAS-CONF-2013-013,
%ATLAS-CONF-2013-013,
CMS-PAS-HIG-13-002,CMS-PAS-HIG-13-004,ATLAS-CONF-2012-161,CMS-PAS-HIG-12-044},
with no observation of any other easily accessible states at the low energies probed so far.
However, the neutrino masses, or more generally the flavor patterns of all the fermions, 
as well as cosmological problems on the nature of dark matter and the origin of matter antimatter asymmetry continue to supply an incentive for extending SM.

Motivated by weakly coupled dark matter candidates (WIMPs) one can extend SM by adding new states which are singlet under the SM gauge interactions. 

The simplest possibility is to consider a real singlet scalar $s$ on top of the SM \cite{McDonald:1993ex}. 
With this model one can attempt to explain the dark matter abundance: if a discrete $Z_2$ symmetry is imposed on the singlet scalar, it is stabilized, and with a suitable mass and weak enough coupling to the SM fields can constitute the observed dark matter abundance \cite{McDonald:1993ex}. Furthermore, with suitable symmetry breaking patterns, the model can also lead to first order electroweak phase transition (EWPT) and abet electroweak baryogenesis \cite{McDonald:1993ey,Cline:2012hg}.
In this simple model the singlet scalar couples to the SM only through the Higgs portal coupling
\be
\lambda_{sh}|\Phi|^2 s^2,
\ee
which determines the dark matter relic density through the annihilation and thermal decoupling. It also determines the strength of the electroweak phase transition. This in turn creates some tension: the portal coupling should be small enough to yield the correct relic density and at the same time large enough to allow for a strong electroweak transition. Due to this tension,  in this model it is difficult to generate both a sizeable fraction of the observed dark matter abundance and a strong enough first order electroweak phase transition simultaneously.

In addition to the low energy phenomenology, extended scalar sectors may have
consequences for the very early universe.
If the scalar fields are light during inflation, they will acquire fluctuations proportional to the inflationary scale $\delta h\sim
\delta s\sim H_{\rm inf.}$. The question then arises: after inflation, how and by which rate do the scalars attain their vacuum values? Is it obvious that by the time of electroweak symmetry breaking, all the fields
have relaxed into the vacua that are requisite for successful baryogenesis and/or observed dark matter abundances?

In this paper we will explore the behaviour of the scalar fields $h$, the physical Higgs, and $s$, the real singlet, in the $Z_2$ model. We assume their energy
densities are subdominant during inflation, $V(h,s)\ll 3 H_{\rm inf}^2$,
and that both the scalar fields are light, i.e. $\partial_h^2V\ll H_{\rm inf}^2$ and
$\partial_s^2V\ll H_{\rm inf}^2$. If one of the fields is effectively
massive initially, the energy density associated with its oscillations
decreases exponentially as a function of the number of e-foldings. The field then
rapidly reaches the regime where its effective mass is small as
compared to the inflationary scale. If on the other hand $V(h,s)\sim
3 H_{\rm inf}^2$ and the kinetic energy densities are negligible,
then either the higgs or the singlet $s$ should act as an inflaton.
Here we will not explore this possibility.

The paper is organized as follows: In section \ref{model} we introduce the scalar sector of the model and outline the
low energy particle physics phenomenology. Then, in section \ref{init} we discuss the initial conditions from inflation. In section \ref{decay} we consider the decay of the scalar condensates and find out the appropriate
decay rates. In section \ref{checkout} we conclude and consider the outlook and possible directions for further work.

\section{The model and low energy particle phenomenology}
\label{model}
%%%%%%%%%%%%%%%%%%%%%%%%%%%%%%%%%%%%%%%%%%%%%%%%%%%%%%%%%%%%%%%%%%%%%%%%%%%%%%%%%%%%%%%%%%%%%%%%%%%%
%%%%%%%%%%%%%%%%%%%%%%%%%%%%%%%%%%%%%%%%%%%%%%%%%%%%%%%%%%%%%%%%%%%%%%%%%%%%%%%%%%%%%%%%%%%%%%%%%%%%
The scalar sector of the model is specified by the potential
\be
V(\Phi,S)=m_h^2\Phi^\dagger\Phi+\lambda_h(\Phi^\dagger\Phi)^2+\frac{1}{2}m_s^2 s^2+\frac{\lambda_s}{4}s^4+\frac{\lambda_{sh}}{2}(\Phi^\dagger\Phi)s^2,
\label{scalarpot}
\ee
where $\Phi$ and $s$ are, respectively, the usual Standard Model Higgs doublet and a real singlet scalar.
We have imposed a $Z_2$ symmetry on the Lagrangian under $s\mapsto -s$.
The Higgs doublet is written in terms of the components as
\be
\Phi=\begin{pmatrix} \phi^+\\ \frac{1}{\sqrt{2}}(h^0+i\eta^0+\nu)\end{pmatrix},
\ee
where the superscript refers to the electric charge of the components.

A strong first-order electroweak transition is a prerequisite for successful electroweak baryogenesis \cite{Kuzmin:1985mm} and it is well known that the electroweak phase transition in the SM is not of first order but a smooth crossover \cite{Kajantie:1996mn,Rummukainen:1998as}. 
If the electroweak sector of the SM was fully perturbative, a first-order phase transition would arise from a cubic term generated in the Higgs effective one-loop potential by the thermal effects of fields coupled to the Higgs. However, addition of a singlet scalar can sufficiently modify the picture already by tree level effects due to the presence of $T$-independent dimensional parameters appearing in the scalar potential and lead to a strong first-order transition. 
Consequently, the ratio $v(T_c)/T_c$ which controls the sphaleron erasure of the baryon asymmetry can be large and lead to successful electroweak baryogenesis. 
At finite temperature the effective potential
is treated at the mean field level, taking into account the finite temperature corrections to the masses:
\be
m_h^2\mapsto m_h^2+c_h T^2, \quad m_s^2\mapsto m_s^2+c_s T^2,
\label{scalarmass}
\ee
where the factors $c_h$ and $c_s$ are
\bea
c_h &=& \frac{1}{12}\left( c_{\rm{SM}}+6\lambda_h+\frac{1}{2}\lambda_{sh}\right),\nonumber \\
c_s &=& \frac{1}{12}\left( 2\lambda_{sh}+3\lambda_s \right)
\label{c_h}
\eea
and we defined
\be
c_{\rm{SM}}=\frac{9}{4}g^2+\frac{3}{4}g^{\prime 2}+3y^{2}_{t}.
\ee

There are some immediate constraints on the parameters of the scalar potential:

The stability of the potential clearly requires
$\lambda_h>0$ and $\lambda_s>0$. Furthermore
\be
\lambda_{sh}>-2\sqrt{\lambda_h\lambda_s}.
\label{couplconstr}
\ee
Note that stability allows $\lambda_{sh}$ to be negative. Third, the masses of the fields around the vacuum $\nu=$ 246 GeV are given by
$M_s^2=m_s^2+\nu^2\lambda_{sh}/2$,
and, as usual, $M_h^2=2 \nu^2\lambda_h$, which gives (with $M_H=126$ GeV) $\lambda_h=0.131$. 

The scalar potential (\ref{scalarpot}) has five parameters. With the input $\nu=246$ GeV and $m_h=126$ GeV three free parameters remain; we take these to be $m_s^2$, $\lambda_s$, and $\lambda_{sh}$. Their values are constrained by the above relations. The couplings of the Higgs field to the SM matter are given by the Yukawa sector of the Standard Model.

The low energy phenomenology of this model has been studied thoroughly in the literature already \cite{McDonald:1993ex,McDonald:1993ey,Cline:2012hg}. Here we now discuss the essential features which emerge
from those studies, namely the vacuum structure and allowed ranges of values of the portal coupling $\lambda_{sh}$. These arise basically from relating 
the constraints from the LHC data, the possibility of the singlet to explain 
the dark matter relic density, and the possibility of achieving a strong electroweak phase transition. 

In order to obtain modifications to the electroweak phase transition at tree level there must be a minimum along the $s$-direction which becomes a global minimum at some intermediate temperature above the electroweak scale.
However, to guarantee the stability of $s$, the global minimum of the potential at $T=0$ must be at $\langle s\rangle=0$.  
This requirement implies that
\be
|m_s^2|\le \nu^2\sqrt{\lambda_h\lambda_{s}},
\label{massconstr}
\ee
where $\nu=246$ GeV is the electroweak scale. The expression of the singlet mass can be alternatively written as 
$m_s^2=M_s^2-\nu^2\lambda_{sh}/2$, and in order for there to be symmetry breaking in the $s$-direction at all, we need to have $m_s^2<0$. 
Equivalently, $\lambda_{sh}>M_s^2/\nu^2\sim 10^{-1}$ for $M_s\sim 50$ GeV, which is a realistic lower bound for the mass of a dark matter particle produced via annihilations as the system departs from equlibrium.

There are different possibilities how the dark matter abundance may arise. If $s$ is a thermal relic,
i.e. its number density is determined by the usual freeze-out calculation \cite{Gondolo:1990dk}. The standard approximate solution to the Lee-Weinberg equation is
\be
\Omega_s h^2\simeq \frac{1.07\cdot 10^9 x_f}{\sqrt{g_\rho}M_{\rm{pl}}\langle v\sigma\rangle},
\ee
where $g_\rho$ is the effective number of thermal degrees of freedom,
$x_f=m_s/T_f\sim 10\dots 20$ and $T_f$ is the freeze-out temperature.
In this model the relevant annihilation channels entering into $\langle v\sigma\rangle$
are $ss\rightarrow hh$, $ss\rightarrow VV$ $(V=W,Z)$ and
$ss\rightarrow \bar{q}q$.

Since the dark matter abundance can consist of several components, we define the
relative relic density $f_{\rm{rel}}=\Omega h^2/0.12$, with the value $(\Omega h^2)_c=0.12$ from Planck \cite{Ade:2013zuv}, 
This allows to consider the subdominant cases also, taking $1\ge f_{\rm{rel}}\ge 0.01$.

The basic constraint arising from the LHC is that if the singlet $s$ is light enough, it provides an invisible decay channel for the Higgs. The decay rate for the Higgs to decay into invisible channel $h\rightarrow ss$ is given by
\be
\Gamma_{\rm{inv}}=\frac{\lambda_{sh}^2 \nu^2 }{32\pi M_h}\sqrt{1-4M_s^2/M_h^2}.
\ee
and the corresponding branching ratio is
\be
{\rm{BR}}_{\rm{inv}}=\frac{\Gamma_{\rm{inv}}}{\Gamma_{\rm{tot,SM}}+\Gamma_{\rm{inv}}}.
\ee
The 2$\sigma$ limit for the invisible decay width from the current LHC data assuming SM-like couplings between the Higgs boson and fermion and gauge fields of SM is ${\rm{BR}}_{\rm{inv}}=0.19$. This gives a constraint on the allowed value of $\lambda_{sh}$, and practically rules out all values of $M_s\lsim M_h/2$ if the singlet $s$ is simultaneously required to contribute at least 1\% to the observed relic density, $f_{\rm{rel}}>0.01$ \cite{Cline:2012hg}.

For typical electroweak scale WIMP, $M_s\sim{\cal O}(100\,{\rm{GeV}})$, producing $f_{\rm{rel}}\simeq 1\dots 0.1$ requires 
$10^{-2}\le |\lambda_{sh}|\le 10^{-1}$  \cite{Cline:2012hg}. As we already derived, the strong EWPT requires $\lambda_{sh}\ge 0.1$. Hence, both strong EWPT and reasonable fraction of the dark matter abundance cannot be simultaneously realized in this model. 

If we concentrate only on the dark matter, then the relic density can also be produced out of equibrium via the decay of the thermal Higgs bosons $h^0\rightarrow ss$ at low temperature \cite{McDonald:2001vt}, also known as the freeze-in mechanism \cite{Hall:2009bx,Yaguna:2011qn,Klasen:2013ypa,Blennow:2013jba}. Analogously to the approximate result in the freeze-out case, the dark matter abundance in this case is given by
\be
\Omega_s h^2\simeq \frac{1.09\cdot 10^{27}}{g_s\sqrt{g_\rho}}\frac{m_S\Gamma_{h^0\rightarrow SS}}{m_h^2},
\ee
where $g_{s,\rho}$ denotes the effective degrees of freedom for the entropy and energy density, respectively. The relation between the 
mass and temperature is different when comparing to the freeze-out mechanism, for which the relic abundance is created at $x=m_s/T\simeq 20\dots 30$ when the system departs from equlibrium: the freeze-in yield arises during the epoch $x\sim 2\dots 5$ \cite{Hall:2009bx}. 

The essential feature of this mechanism is that the coupling required for the production of sufficient relic abundance is superweak; one obtains a parametric estimate
\be
\lambda_{sh}\simeq 10^{-11} \left(\frac{\Omega_sh^2}{0.12}\right)^{1/2}\left(\frac{{\rm{GeV}}}{M_s}\right)^{1/2}.
\ee
Such weak couplings are not at odds with LHC data or the current direct search limits, and allows the dark matter particle also to be light. 

To summarise: with a simple extension, Eq. (\ref{scalarpot}), of the Standard Model scalar sector, one can realize different phenomenological goals at low energies. If the portal coupling is large enough  $\lambda_{sh}\simeq 0.1$, the model
may lead to a strong EWPT required by any successful baryogenesis scenario. On the other hand, creating (a sizeable fraction of) the observed dark matter abundance is possible via freeze-out or freeze-in mechanisms corresponding to the parametric regions for the portal coupling $\lambda_{sh}\simeq 10^{-2}$ and $10^{-11}$, respectively. The value of the portal coupling in the case of freeze-in can be shifted by few orders of magnitude up by considering lighter dark matter, or down by considering heavier or subdominant dark matter candidate.

\section{Initial conditions set by inflation}
\label{init}

If the scalar fields are light during inflation, the mean fields will acquire
fluctuations proportional to the inflationary scale $\delta h\sim
\delta s\sim H_{\rm inf}$. We assume $H_{\rm inf}\gg T_{\rm EW}$
so that the quadratic terms in the potential (\ref{scalarpot}) can
be neglected in investigating the dynamics during inflation. The
potential then reduces to the form

  \be
  \label{Vinf}
  V(h,s)=\frac{\lambda_h}{4}h^4+\frac{\lambda_s}{4}s^4+\frac{\lambda_{sh}}{2}h^2s^2
  .
  \ee

We will explore the behaviour of the fields assuming their energy
density is subdominant during inflation $V(h,s)\ll 3 H_{\rm inf}^2$
and that the fields are light, i.e. $V_{hh}\ll H_{\rm inf}^2$ and
$V_{ss}\ll H_{\rm inf}^2$. If one of the fields is effectively
massive initially, the energy density associated to its oscillations
decreases exponentially in the number of e-foldings. The field then
rapidly reaches the regime where its effective mass is small
compared to the inflationary scale. If on the other hand $V(h,s)\sim
3 H_{\rm inf}^2$ and the kinetic energy densities are negligible
then either the Higgs or the singlet $s$ should act as an inflaton.
In this work we will not explore this possibility.

The dynamics of the subdominant light fields can be investigated
using the stochastic approach \cite{Starobinsky:1994bd}. The
average behaviour of the fields on superhorizon scales is then
controlled by the Fokker-Planck equation for their distribution
$P(h,s)$
  \be
  \label{fokker-planck}
  \frac{\partial P(h,s)}{\partial t} = \sum\left[\frac{H^3}{8\pi^2}\frac{\partial^2 P(h,s)}{\partial \phi^2}+
  \frac{1}{3H}\frac{\partial}{\partial \phi}\left(P(h,s)
  \frac{\partial V(h,s)}{\partial \phi}\right)\right]~,
  \ee
where the sum runs over $\phi=h,s$.
Here we have assumed the Hubble rate during inflation to be a
constant $H=H_{\rm inf}$. This amounts to neglecting slow roll
suppressed correction terms. As the system reaches equilibrium, the
distribution becomes time independent $\partial_t P_{\rm
eq.}(h,s)=0$. It can easily be seen that a solution for the
equilibrium distribution is given by
  \be
  P_{\rm eq.}(h,s)=N {\rm exp}\left(-\frac{8\pi^2 V(h,s)}{3H^4}\right)\
  ~,
  \ee
where $N$ is the normalization constant.

Assuming the coupling $\lambda_{sh}$ in (\ref{Vinf}) is small we can
then straightforwardly work out the desired equilibrium correlation
functions as a perturbative expansion in $\lambda_{sh}$. To first
order in the coupling we find
  \bea
  \langle h^{2m} s^{2n}\rangle&=&
  \frac{H^{2m+2n}}{\lambda_h^{m/2}\lambda_s^{n/2}}\left(\frac{3}{2\pi^2}\right)^{(m+n)/2}
  \frac{\Gamma\left(\frac{1}{4}+\frac{m}{2}\right)\Gamma\left(\frac{1}{4}+\frac{n}{2}\right)}
  {\Gamma\left(\frac{1}{4}\right)^2}\\\nonumber
  &&\times\left(1+\frac{2\lambda_{sh}}{\sqrt{\lambda_h\lambda_s}}
  \left(\frac{\Gamma\left(\frac{3}{4}\right)^2}
  {\Gamma\left(\frac{1}{4}\right)^2}-\frac{\Gamma\left(\frac{3}{4}+\frac{m}{2}\right)\Gamma\left(\frac{3}{4}+\frac{n}{2}\right)}
  {\Gamma\left(\frac{1}{4}+\frac{m}{2}\right)\Gamma\left(\frac{1}{4}+\frac{n}{2}\right)}\right)\right)\
  .
  \eea
In particular, the variance $\langle h^2 \rangle $ to this order is
given by
  \bea
  \langle h^2 \rangle &=&
  \frac{H^2}{\sqrt{\lambda_h}}\sqrt{\frac{3}{2\pi^2}}
  \frac{\Gamma\left(\frac{3}{4}\right)}{\Gamma\left(\frac{1}{4}\right)}
  \left(1+\frac{2\lambda_{sh}}{\sqrt{\lambda_h\lambda_s}}
  \left(\frac{\Gamma\left(\frac{3}{4}\right)^2}{\Gamma\left(\frac{1}{4}\right)^2}-\frac{1}{4}\right)\right)\\\nonumber
  &\simeq&
  \frac{0.132
  H^2}{\sqrt{\lambda_h}}\left(1-\frac{0.272\lambda_{sh}}{\sqrt{\lambda_h\lambda_s}}\right)
  \ ,
  \eea
and analogously for $\langle s^2 \rangle$.

The root mean square values $h_{*} = \sqrt{\langle h^2\rangle}$
characterize the typical magnitudes of the scalar condensates
generated during inflation. If the coupling between the Higgs and
singlet is small $\lambda_{sh}\ll \sqrt{\lambda_h\lambda_s}$ we thus
arrive at the results
  \be
  h_{*}={\cal O}(0.1) \frac{H}{\lambda_h^{1/4}}\ ,\qquad s_{*}={\cal O}(0.1)
  \frac{H}{\lambda_s^{1/4}}\ .
  \label{h,s_*}
  \ee
In the following we will take these results as inflationary predictions for the initial values of the scalar condensates at the end of inflation. This sets the starting point for our discussion of the subsequent cosmological evolution.

%%%%%%%%%%%%%%%%%%%%%%%%%%%%%%%%%%%%%%%%%%%%%%%%%%%%%%%%%%%%%%%%%%%%%%%%%%%%%%%%%%%%%%%%%%

\section{Decay of the scalar condensates}
\label{decay}

At the onset of the hot big bang epoch after the end of inflation
the Higgs field and the singlet are displaced far away from their
vacuum values. The rate of relaxation towards the vacuum stage plays
a crucial role for the cosmological implications of the scenario.
For example, whether it is possible to obtain a first order
electroweak transition depends crucially on the field values at the
electroweak scale.

Soon after the end of inflation the condensates (\ref{h,s_*}) become
massive and start to oscillate around the minimum of the potential
with the amplitude diluted by the expansion of space. The
condensates will also decay into themselves and other particles, the
Higgs directly through its couplings to SM fields and the singlet
through Higgs mediated processes. Several decay channels remain
kinematically blocked for a long time due to large thermal
corrections and background field values. The thermal corrections
will also block non-perturbative decay channels which otherwise
would play a significant role in the dynamics. Here we will investigate in detail the decay of the scalar condensates and determine the time scale for their relaxation to vacuum stage. For other recent studies of thermal dissipation rates in scalar field theory, see \cite{Mukaida:2012qn, Mukaida:2013xxa, Mukaida:2014yia}.

\subsection{Background dynamics}

We assume the inflaton field(s) decay instantaneously into SM
particles at the end of inflation so that the reheat temperature is
directly given by the inflationary scale
  \beq
  \label{instantRH}
  T_{*}=\left(\frac{90}{\pi^2 g_*}\right)^{1/4} \left(H_{*}M_{P}\right)^{1/2}
  \simeq 8 \times 10^{15} {\rm GeV}\left(\frac{r}{0.1}\right)^{1/4}
  \ .
  \eeq
Here $r$ is the tensor-to-scalar ratio; if the BICEP2 detection of the B-modes were to pass scrutiny, $r \simeq 0.1$ \cite{Ade:2014xna}. For our purposes, the precise
value of the tensor-to-scalar ratio is not essential.
The universe enters a radiation dominated stage with $a\propto
t^{1/2}$.

Interactions with the thermal bath very quickly generate a thermal
mass for the Higgs condensate. The thermal Higgs mass
(\ref{scalarmass}) completely dominates over the zero temperature
part (\ref{h,s_*})

  \beq
  m_{h}^2(T) \simeq c_h T_{*}^2 \sim 3\times 10^{4} \frac{c_h}{\sqrt{\lambda_h}}
  \left(\frac{r}{0.1}\right)^{-1/2}
  m_{h}^2\ .
  \eeq

Ignoring the decay processes, the condensate would then obey the
equation of motion
  \beq
  \label{higgseom}
  \ddot{h}+\frac{3}{2t}\dot{h}+c_h T^2 h = 0\ ,
  \eeq
which gives for the envelope $\bar{h}(T)$ the result
  \beq
  \label{higgsenvelope}
  \bar{h}(T)\simeq 7\times 10^{-4}
  \left(\frac{r}{0.1}\right)^{1/2}\left(\frac{0.01}{\lambda_h}\right)^{1/4}
  T\ .
  \eeq

The singlet on the other hand interacts with the thermal bath only
through its coupling to the Higgs field and with the rate\footnote{Strictly speaking this rate holds true only for a process $hh\rightarrow ss$ due to uneven number densities between Higgs and singlet particles but it can nevertheless be used to determine the temperature scale at which the singlet field finally succeeds to maintain the thermal corrections. This should not deviate much from the real rate because the Higgs particle abundance after inflaton decay, and thus also the singlet particle production after $\Gamma_{hh\rightarrow ss}/H > 1$ is reached, are assumed to be very high.}

  \beq
  \label{thermalsinglet}
  \frac{\Gamma_{ss\rightarrow hh}}{H} \sim 10^{-2} \lambda_{sh}^2 \frac{T}{H} \sim \lambda_{sh}^2 \frac{10^{16}{\rm
  GeV}}{T}\ .
  \eeq
If $\lambda_{sh} \ll 1$, the scattering rate is negligible at the onset of the
hot big bang epoch and the singlet starts to feel the thermal bath
only after the temperature has decreased below $T_{\rm s}= \lambda_{sh}^2
10^{16}$ GeV. This happens above $T_{EW}$ only if $\lambda_{sh}\gtrsim 10^{-7}$.

Ignoring again all the decay processes we would then obtain the equation of motion
  \beq
  \label{singleteom}
  \ddot{s}+\frac{3}{2t}\dot{s} = -\Big\{{\lambda_s s^3\ ,\qquad T > \lambda_{sh}^2
10^{16}{\rm  GeV} \atop c_s T^2 s\ ,\qquad T < \lambda_{sh}^2
10^{16}{\rm GeV}}\
  .
  \eeq
In both cases the solution for the envelope is given by the same
expression
  \beq
  \label{singletenvelope}
  \bar{s}(T) \simeq 5\times 10^{-3} \lambda_s^{-3/8}
  \left(\frac{r}{0.1}\right)^{1/4} T\ .
  \eeq
We will now move on to study how these solutions change as a result of the decay processes of the oscillating condensates.

\subsection{Decay of the Higgs}

We start by considering non-perturbative decay channels. In non-perturbative particle production ordinary SM and singlet particles are produced by a resonant decay of the Higgs condensate. The efficiency of this process depends both on the oscillation frequency and amplitude of the decaying field and is usually characterised by a resonance parameter $q$ referring either to an efficient broad resonance ($q>1$) or usually inefficient narrow resonance ($q<1$). An efficient particle production also requires the field to evolve non-adiabatically \cite{Kofman:1997yn,Greene:1997fu}.

For example, at subhorizon scales the equation of motion of the Higgs particles can be written in the form

  \be
  \label{higgsmode}
  \ddot{h}_k +\left(\omega^2_k(t) \equiv \frac{k^2}{a^2}+\lambda_h h^2(t)+\lambda_{sh}s^2(t)+c_hT^2\right) h_k = 0 ,
  \ee
where $h(t)$ and $s(t)$ are given by (\ref{higgseom}) and (\ref{singleteom}), respectively. Note that this makes the effective frequency $\omega_k(t)$ conformally invariant.
  
As the oscillation frequency of the Higgs condensate is proportional to the high temperature (\ref{higgseom}), all resonances are kept narrow, $q=\lambda_{h}h^2_*/(4c_hT_*^2)\ll1$, and the condensate evolves adiabatically, $\dot{\omega}_k(t)/\omega^2_k(t)<1$.

Although the narrow resonance regime does not necessarily require violation of the adiabaticity condition, in our case the first resonance band, which usually is the most efficient one, is blocked from the beginning and the resonance in the second band quickly terminates by the backreaction of induced particle production \citep{Kofman:1997yn}. Thus this effect cannot compete with the perturbative effects (see section \ref{higgstwoloop}). The same conclusion holds also for the production of all the other SM and singlet particles. We thus conclude that all effective non-perturbative decay channels of the Higgs condensate are subdominant at least until $T\sim m_{top}$.

\subsubsection{Perturbative decay at one-loop level}
The Standard Model Higgs could decay perturbatively into quarks, leptons and gauge bosons. In earlier work \cite{Enqvist:2013kaa} it was shown that at $T=0$ either the decay of the condensate is very inefficient or the most effective decay channels are kinematically blocked for many Hubble times.

However, due to interactions of the condensate with the thermal background and unusual dispersion relations the conditions for perturbative decay are not as straightforward as in zero temperature. In this section we study numerically the threshold conditions for all perturbative decay channels by using the imaginary part of the two-point functions computed at one-loop level. The thermal masses and threshold conditions for all the fields, including scalars, fermions and gauge fields used in the calculation can be found e.g. in \cite{Elmfors:1993re}.

\subsubsection*{Fermions}
In thermal bath, there appear two kinds of fermionic excitations, called particles and holes, which both have dispersion relations drastically changed from the $T=0$ case \cite{Elmfors:1993re}. At one-loop level there are two possible decay channels for a scalar interacting with fermionic excitations: decay and absorption. The inverses of these processes are also possible.

By performing a numerical computation we find both channels to be kinematically blocked at all energy scales up to $H=10^{14}$ GeV. Thus there is no Higgs decay into fermions at one-loop level.

\subsubsection*{Gauge fields}
\label{gaugefields}

When the Higgs condensate is hit by a Higgs particle from the thermal bath it can annihilate to produce a gauge boson. We study only the longitudinal bosons because for a Higgs condensate at rest ($p_0=m_h$, $\bf{p}=0$), only the longitudinal part contributes to the thermal decay of the Higgs field due to the derivative coupling between the Higgs and gauge fields.

The decay was again studied numerically. We find that the absorption channel is kinematically blocked at all energy scales up to $H=10^{14}$ GeV. Thus there is no Higgs decay into gauge bosons at one-loop level.

\subsubsection*{One-leg-in-vacuum diagrams}
\label{VEVdecay}

While the Higgs keeps oscillating, there is an effective, non-zero background field, which generates three-point vertices. By expanding the Higgs around the envelope field value $\bar{h}(T)$ (\ref{higgsenvelope}), $h\rightarrow \bar{h}(T)+h$, one obtains the so-called one-leg-in-vacuum diagrams where the magnitude of the coupling is modified by the field value. These diagrams, including Higgs condensate decay and several different absorption processes, provide an another thermal decay channel for a Higgs condensate.

Only such diagrams which contain loops represent physically relevant processes, meaning that at this level there is no Higgs decay into fermions due to their Yukawa-type coupling. However, this time also the Higgs decay to transversal bosons is allowed.

We stress that in order to have these processes well defined, the induced decay rate must be larger than the rate of change in the background field value. The decay rate is suppressed by the high temperature,

\be
\Gamma_{h} \simeq \frac{g^4\bar{h}^2(t)}{m_W(T)} \sim \frac{g^3}{\sqrt{\lambda_h}}\frac{H^2}{T},
\ee
whereas the rate of change in the background field value is roughly

\be
\left|\frac{\dot{h}(t)}{h(t)}\right| \simeq \sqrt{c_h}T ,
\ee
following from the full solution of (\ref{higgseom}). Therefore we have

\be
\frac{\Gamma_{h}}{\left| \dot{h}(t)/h(t)\right|} \simeq \frac{g^2}{\sqrt{\lambda_h}}\left(\frac{H}{T}\right)^2 \ll 1 
\ee
nearly everywhere, which has also been verified numerically by taking into account the oscillating phase in $h(t)$.

This means that our formalism is not applicable for studying these processes in a wildly oscillating condensate background. A more detailed investigation of this channel would require a lattice simulation, although the suppression by high temperature keeps this channel subdominant. Therefore we do not expect that these one-leg-in-vacuum processes would give any significant contribution to the decay of Higgs condensate whenever a thermal background is present.

\subsubsection{Perturbative decay at two-loop level}
\label{higgstwoloop}

At two-loop and higher level there are no energy thresholds and there is no   kinematical obstruction for the decay of a scalar field. The magnitude of two-loop processes can be estimated by calculating the "rising-sun" diagrams which typically give the dominant contribution.

The bosonic "rising-sun" diagram gives for the thermalization rate \cite{Elmfors:1993re}

\be
\Gamma_{h} = \frac{3}{256\pi}\frac{g^4}{m_h(T)}T^2.
\ee
This is the dominant decay channel for the Higgs condensate.

By requiring $\Gamma_{h}/H\geq1$, we obtain the thermalization to begin at

\be
t\gtrsim 0.5\times g^{-8}\left(\frac{r}{0.1}\right)^{1/2}H_0^{-1}\simeq 140H_0^{-1} ,
\ee
corresponding to $T\simeq 10^{-2}T_*$. This leads to a situation where the condensate relaxes down to its minimum very fast, meaning that the field is found at $\langle h \rangle =0$ well above $T_{EW}$.

\subsection{Decay of the singlet}

Let us first make a rough estimate of the singlet field value at the electroweak scale by taking into account only the dilution of the condensate due to the expansion of the universe.

The singlet scales as $s\propto a^{-1}$ both when $V\sim \lambda_s s^4$ and $V\sim s^2T^2$. The thermal mass becomes dominant in the evolution as the temperature has decreased below $T_{\rm s}= \lambda_{sh}^2 10^{16}$ GeV (\ref{thermalsinglet}) which can easily correspond to a temperature above the electroweak scale. 

The thermal mass becomes comparable to the zero temperature mass at

  \be
  \label{singletmass}
  T_{\rm eq}^2 = \frac{m_s^2}{c_s}\ ,
  \ee
and the corresponding field value in units of $m_s$ is given by 
  \be
  \frac{s_{\rm eq}}{m_{s}} \simeq 0.1
  \left(\frac{\lambda_s}{c_s}\right)^{1/2}\left(\frac{\lambda_s}{0.01}\right)^{-3/4}
  \left(\frac{r}{0.1}\right)^{1/4}\ .
  \ee
The singlet has therefore essentially rolled down to zero as it
starts to feel the structure of the underlying $T=0$ part of the
potential. If $m_s^2 < 0$, the potential is unstable and the singlet
will rapidly fall from $s_{\rm eq}$ to its vacuum value $s_0=
|m_s|/\sqrt{\lambda}$. This happens in a time scale $t\sim 1/|m_s|
\ll H^{-1}$. Therefore, the singlet reaches its vacuum value at
around $T\sim T_{\rm eq}$.

\subsubsection{Non-perturbative decay}
\label{singletnonpertdec}

The above conclusion considered the dilution of the singlet condensate only. The relaxation of the singlet down to its minimum can be eased by allowing also decay of the condensate.

The singlet condensate can decay non-perturbatively both into Higgs and singlet particles. For this, we have to cases: the one in which the singlet has not yet acquired a large thermal mass and the one in which it has. The transition between these two cases happens roughly at $T_{\rm s}= \lambda_{sh}^2 10^{16}$ GeV (\ref{thermalsinglet}). 

Let us first study the case where the singlet has already acquired a thermal mass. Again we find that the resonance is always very narrow, $q\ll1$, and the condensate evolves adiabatically. However, the induced particle production backreacts to the resonance and quickly terminates it. Thus this effect is negligible. 

Let us then turn into the case where the singlet has not yet acquired a thermal mass. The decay rate for the singlet condensate decaying into Higgs particles has been calculated to be \cite{Ichikawa:2008ne}

\be
\label{ichikawarate}
 \Gamma_{s} = C_2\frac{\lambda^2_{sh}}{8\pi m^3_s}\rho_s ,
  \ee
where $C_2\approx 8.86$ is a numerical constant and $\rho_s \simeq \frac{\lambda_s}{4}s^4$ is the energy density of the condensate.

The decay rate is valid down to $T_{EW}$ if the singlet never acquires a thermal mass. This is true if $\lambda_{sh}<10^{-7}$ (\ref{thermalsinglet}), in which case we see from (\ref{ichikawarate}) that the singlet condensate decay into Higgs particles is completely negligible effect.

There is also an another option: when the Higgs condensate has decayed, the equation of motion for the singlet particles is the conformally invariant Lame equation

\be
  s''_k(\eta) +  \left( \frac{k^2}{\lambda_ss^2_*} + 3 \text{cn}^2\left(\sqrt{\lambda_ss^2_*}\eta , \frac{1}{\sqrt{2}}\right)\right) s_k = 0 ,
  \ee
where $\eta$ is the conformal time and cn is the Jacobi elliptic cosine, obtained from the full solution of (\ref{singleteom}). With the prime ' we denote the derivative with respect to the conformal time. In this case the resonance is broad, $q=3$, and an efficient particle production can take place but only for a short time due to the backreaction of created particles. The rapid restructuring of the resonance band shuts the resonance down before the energy density of the singlet particles constitutes more than $0.2\%$ of the total energy in the condensate \cite{Greene:1997fu}. Thus also this fragmentation effect is negligible.

The conclusion therefore is: as long as the singlet never acquires a thermal mass before $T_{EW}$, it never thermalizes.

\subsubsection{Perturbative decay at one-loop level}

Because there is no tree-level decay of the singlet condensate, the one-leg-in-vacuum diagrams are the only possible decay channels below the two-loop level. There are two possible processes: the singlet's one-leg-in-vacuum decay process, $s(\bar{p}=0)\rightarrow h+h$, and the singlet absorption process, $s(\bar{p}=0)+s\rightarrow h$, where the singlet condensate is hit by a singlet particle.

However, as discussed in section \ref{VEVdecay}, the induced decay rate must be larger than the rate of change in the background field value. This is again not the case for the singlet condensate oscillating either in $T^2s^2$ or $s^4$ background.

\subsubsection{Perturbative decay at two-loop level}
\label{singlettwoloop}

For the singlet one finds \cite{Enqvist:2013gwf}

\be
\label{singletrate}
\Gamma_{s,2} = \frac{1}{576\pi}\frac{\lambda^2_{sh}}{m_s(T)}T^2 ,
\ee
corresponding to $T=10^{-10}T_* \gg T_{EW}$ with coupling values $\lambda_{sh}=10^{-4}$, $\lambda_s=10^{-2}$. This is the dominant decay channel for the singlet condensate. We see that by taking the decay into account, the singlet typically relaxes down to its minimum well above $T_{EW}$.

One might worry that the full thermal correction to the singlet mass used in derivation of (\ref{singletrate}) has not yet developed at that point. We have checked that this is the case only if $\lambda_s \lesssim 10^{-3}$. If this is to happen, then the decay via two-loop channel starts immediately after the thermal background has been produced. This happens above $T_{EW}$ as long as $\lambda_{sh}\gtrsim 10^{-7}$ (\ref{thermalsinglet}).

\subsection{Field values at the electroweak scale}

As we have demonstrated above, the Higgs condensate generated during inflation decays into Standard Model particles around the scale $T=10^{-2}T_*$. The Higgs field therefore relaxes to its vacuum configuration well before the electroweak symmetry breaks.

Accounting only for the dilution of the singlet one finds that the behaviour $s\propto a^{-1}$ continues until
  \beq
  \frac{T}{T_{\rm EW}} \simeq 20 \left(\frac{0.01}{\lambda_s}\right)^{1/2}\left(\frac{-m_s^2}{T_{EW}^2}\right)^{1/2} ,
  \eeq
when the negative bare mass term takes over. After this point the singlet falls exponentially fast into its vacuum value. Accounting also for the thermal decay of the singlet, the singlet typically reaches its vacuum configuration well above $T_{EW}$. 

\begin{figure}[htb]
\begin{minipage}[b]{0.50\linewidth}
\centering
\includegraphics[width=\textwidth]{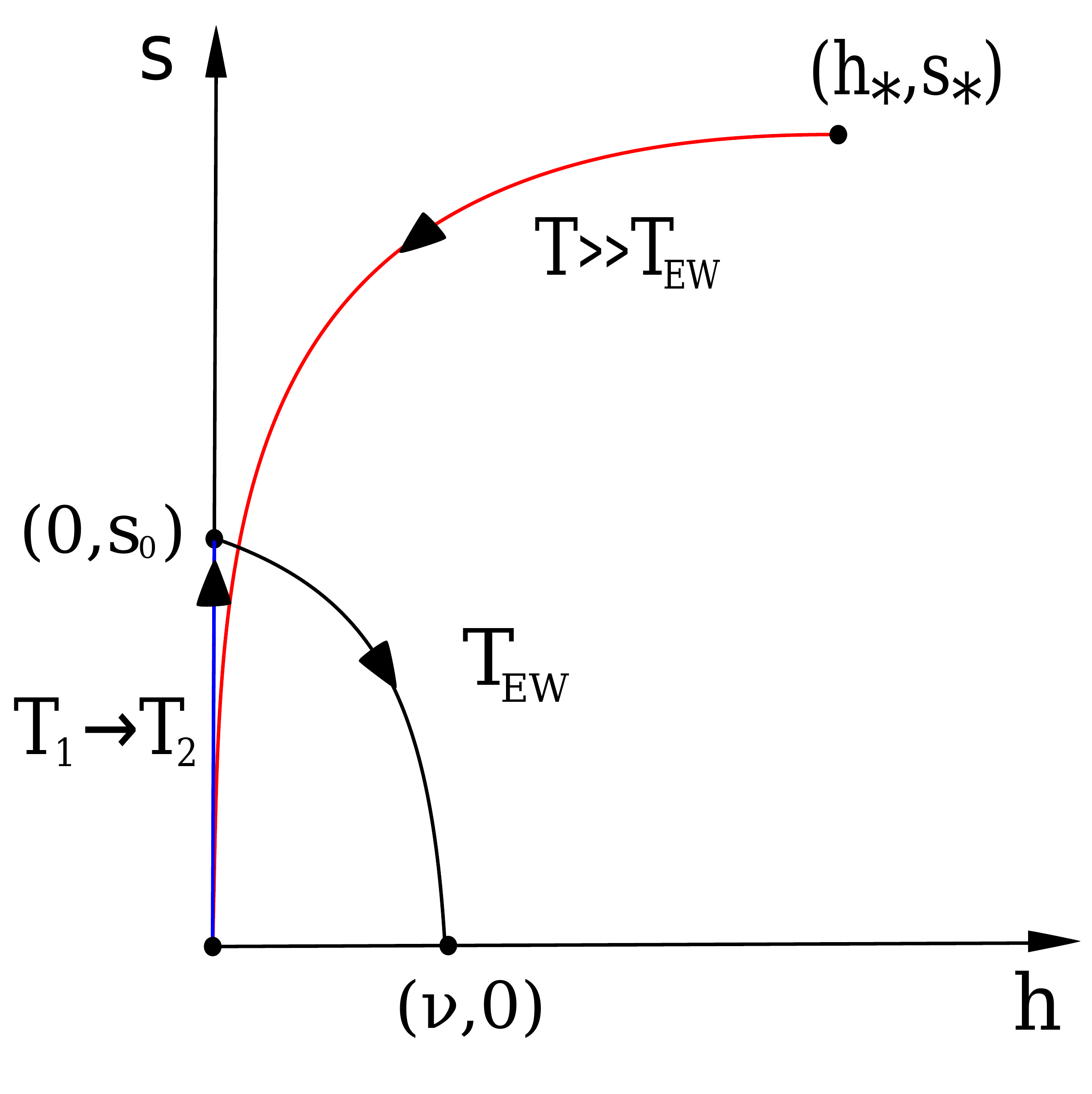}
\end{minipage}
\hspace{0.2cm}
\begin{minipage}[b]{0.50\linewidth}
\centering
\includegraphics[width=\textwidth]{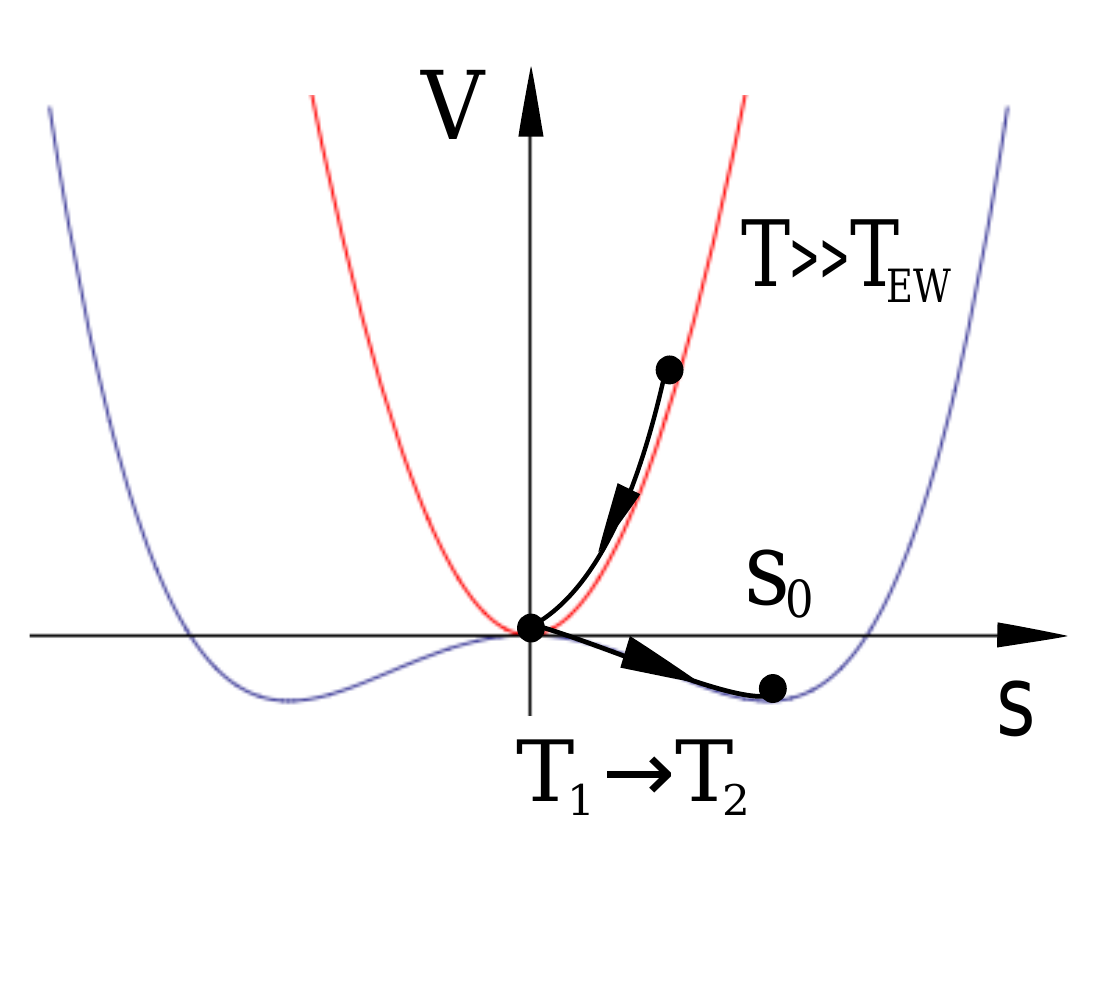}
\end{minipage}
\caption{Left panel: A schematic representation of the field behaviour in $(h,s)$ space. Both fields start at high values, $s_*, h_*\sim H$, but the quick decay of the Higgs condensate drives it down to its non-zero temperature minimum $h=0$ well before $T=T_{EW}$. $T_1$ is defined as a temperature at which the singlet starts to feel its underlying zero temperature mass, whereas at $T_2$ the singlet reaches its non-zero temperature minimum $s_0$. This happens in a time scale $t \ll H^{-1}$, so $T_2\sim T_1$. $T_{EW}$ is the temperature at electroweak symmetry breaking, defined as a temperature at which the two potential minima are equal. At $T_{EW}$ the fields reach their true zero temperature minimum $(\nu,0)$.
%\linebreak
Right panel: A schematic representation of the singlet potential at $T\gg T_{EW}$ (red curve) and at $T \gtrsim T_{EW}$, when the singlet reaches its non-zero temperature minimum $s_0$ at $T_2$ (blue curve).}
\label{fieldspace}
\end{figure}

We thus conclude that also the singlet sector will have relaxed to its vacuum stage sufficient to the baryogenesis mechanism of \cite{Cline:2012hg} by the electroweak scale. A schematic representation of the field behaviour is presented in Fig. \ref{fieldspace}.

\section{Conclusions and Outlook}
\label{checkout}

If the Standard Model remains valid during inflation the Higgs
is a light spectator field \cite{Degrassi:2012ry, Bezrukov:2012sa}, 
\cite{Espinosa:2007qp, Enqvist:2013kaa, Kobakhidze:2013tn, Enqvist:2014bua} 
and inflation should be driven by new physics beyond the Standard Model. Being a light field 
the Higgs gets displaced from its vacuum and the inflationary stage generates
a Higgs condensate with the typical magnitude $h_*  \sim r_T^{1/2}  10^{14}$ GeV. 
While the condensate plays little dynamical role
during inflation, its existence and decay could have significant
impacts on the subsequent hot big bang epoch \cite{Enqvist:2013kaa}

In this work we have concentrated on an extension of the Standard Model where 
a scalar singlet is coupled to the Higgs sector through a portal term $V\supset \lambda_{sh} h^2s^2 $. The singlet could render the crossover electroweak transition
of pure SM \cite{Kajantie:1996mn} into a first order phase transition allowing
for baryogenesis \cite{McDonald:1993ey,Cline:2012hg}. It could also act as dark matter
\cite{McDonald:1993ex,McDonald:1993ey,Cline:2012hg} making portal couplings testable both by cosmological
observations and by laboratory experiments. Assuming renormalizable
couplings and sub-Planckian field values, we have shown that the singlet
is a light spectator field during inflation in analogue to the
Higgs. It is then not enough to concentrate only on predictions
computed in the vacuum state but it is necessary to ask how the
system evolves from the non-vacuum initial conditions set by
inflation.

We have systematically investigated the decay of the
Higgs and singlet condensates generated by inflation. While the
Higgs couplings to other SM fields are large, its decay is delayed by thermal blockings, assuming an instant reheating and a high inflationary scale  $T_{\rm reh}\sim 10^{16}$ GeV as
suggested by the claimed detection of gravitational waves by BICEP2 \cite{Ade:2014xna}. The dominant decay channels for the Higgs condensate are then interactions with the SU(2) gauge bosons at two loop level which  thermalize the Higgs condensate away at $T_{H} \sim 10^{14}$ GeV.  All physical processes below this temperature are therefore insensitive for the non-vacuum initial conditions of the Higgs field.  

The thermalization rate of the singlet is controlled by the magnitude of the portal coupling $\lambda_{sh}$. For $\lambda_{sh}\gtrsim 10^{-7}$ the singlet will thermalize before the electroweak transition. For example, taking the values $\lambda_{sh}=10^{-4}$  and $\lambda_{s} = 0.01$ for the singlet self-coupling as representative examples, we find the thermalization scale given by $T_S \sim 10^{6}$ GeV. On the other hand, for $\lambda_{sh}\lesssim 10^{-7}$ the singlet will not have thermalized by the electroweak scale but both started to produce singlet particles through non-perturbative decay and diluted due to the expansion of space. The dilution is sufficient to bring the singlet close to its vacuum configuration by the electroweak transition and the baryogenesis mechanism of \cite{Cline:2012hg} is therefore not hampered by the non-vacuum initial conditions generated by inflation. However, the out-of-vacuum state of the singlet field could affect physics before the electroweak transition such as the generation of singlet dark matter through a freeze-in mechanism \cite{Hall:2009bx}. It would be interesting to address this question in more detail. Indeed, the dependence of the singlet dark matter on the inflationary scale would constitute a very interesting example of a system where new physics with a tiny coupling to Standard Model could be constrained by carefully investigating its dynamics both during and after inflation. 

\section*{Acknowledgements}
This work was financially supported by the Academy of Finland, projects
1263714 and 1218322 (KE), 257532 (SN), 1257989 and 1263714 (TT) and 267842
(KT).

\bibliography{smsinglet.bib}

\end{document}